\documentclass[prl,twocolumn,preprintnumbers,amsmath,amssymb,superscriptaddress]{revtex4}
\usepackage{graphicx}
\usepackage{latexsym}
\usepackage{amsmath}
\usepackage{graphics}
\usepackage{amssymb}
\usepackage{layout}
\usepackage{verbatim}
\usepackage{amsfonts,epsfig}
\newcommand{\ket}[1]{\left|#1\right>}
\newcommand{\bra}[1]{\left< #1 \right|}
\newcommand{\beq}{\begin{equation}}
\newcommand{\eeq}{\end{equation}}
\newcommand{\HH}{\hat{H}}
\newcommand{\mean}[1]{\langle{#1}\rangle{}}

\begin{document}

\title{Electron Spin Dephasing due to Hyperfine Interactions with a Nuclear Spin Bath}
\author{{\L}ukasz Cywi{\'n}ski} 
\affiliation{Condensed Matter Theory Center, Department of Physics, University of Maryland, College Park, MD 20742-4111, USA}
\author{Wayne M. Witzel}
\affiliation{Naval Research Laboratory, Washington, DC 20375, USA}
\author{S. Das Sarma}
\affiliation{Condensed Matter Theory Center, Department of Physics, University of Maryland, College Park, MD 20742-4111, USA}
\date{\today }

\begin{abstract}
We investigate pure dephasing decoherence (free induction decay and spin echo) of a spin qubit interacting with a nuclear spin bath. While for infinite magnetic field $B$ the only decoherence mechanism is spectral diffusion due to dipolar flip-flops of nuclear spins, with decreasing $B$ the hyperfine-mediated interactions between the nuclear spins become important. We give a theory of decoherence due to these interactions which takes advantage of their long-range nature.
For a thermal uncorrelated bath we show that our theory is applicable down to $B\!\sim$10 mT, allowing for comparison with recent experiments in GaAs quantum dots.
\end{abstract}

\maketitle

\textit{Introduction}.--- 
How a localized electron in a solid loses its spin coherence is one of the oldest problems in condensed matter physics still attracting attention. It has taken on considerable recent significance due to the serious  efforts in attempting to build a scalable quantum computer using electron spins in semiconductors as qubits \cite{Loss_Hanson}.
It is therefore extremely important to understand all aspects of electron spin decoherence in solids qualitatively and quantitatively.
The hyperfine (hf) interaction of the electron spin $S$ with a bath of nuclear spins $J_{i}$ (spin-bath) is the main source of the decoherence of spin qubits at low temperatures, when relaxation effects due to phonons can be neglected. Many theoretical studies have been devoted to this problem \cite{Khaetskii_Coish,deSousa_PRB03,Witzel,Yao,Saikin_PRB07,Deng_PRB06,Coish_PRB08}, and the Spin Echo (SE) decoherence time $T_{\text{SE}}$ has been recently measured in singlet-triplet \cite{Petta_Science05} and single spin qubits \cite{Koppens_PRL08} based on gated GaAs quantum dots. The hf interaction alone can lead to spin decoherence, either through the direct electron-nuclear (e-n) spin flip \cite{Khaetskii_Coish}, or through the effective (mediated by virtual e-n spin flips) intra-bath interaction, which leads to fluctuations of the Overhauser field felt by the electron through the $S^{z}J^{z}_{i}$ part of the hf interaction \cite{Deng_PRB06,Yao,Saikin_PRB07,Coish_PRB08}. These processes are suppressed at large magnetic fields $B$, and above a certain value of $B$ one has to consider the decoherence due to the intrinsic bath dynamics caused by the dipolar interaction between the nuclear spins \cite{deSousa_PRB03,Witzel,Yao,Saikin_PRB07}, the so-called spectral diffusion (SD). 

Presently available theories assume the high magnetic field limit of $\Omega/\mathcal{A} \! \gg \! 1$, where $\mathcal{A}$ is the total hf interaction energy (sum of the hf couplings with all the nuclei) and $\Omega$ is the electron spin splitting. This is fulfilled in SE experiments on electrons bound to the phosphorus donors in Si \cite{Tyryshkin_JPC06}, where the SD theories have shown excellent agreement with observations \cite{Witzel,Saikin_PRB07,Witzel_AHF_PRB07}.
However, in gated GaAs dots this limit corresponds to $B \! \gg \! 1$ T, while the experiments have been done at $B \! \approx \! 10-100$ mT and therefore the existing theories are of dubious validity in this case. The observed $T_{\text{SE}} \! \lesssim \! 1$ $\mu$s in dots with $N\! \sim \! 10^{6}$ nuclei is at least an order of magnitude shorter than the theoretical predictions for SD \cite{Witzel,Yao}. 
This shows that the SE decay in the low field regime in GaAs ($\Omega/\mathcal{A} \!  < \! 1$) is \emph{not} due to the spectral diffusion. 
The existing analytical theory addressing the SE decoherence due to hf interactions only \cite{Yao} gives negligibly small decay at high $B$.

In this Letter we derive a theory of quantum dephasing due to the hf interaction, which applies at low magnetic fields.
Although $\mathcal{A}/\Omega\!  \ll \! 1$ is a sufficient condition for treating the e-n spin flip as a virtual process  \cite{Shenvi_bounds_PRB05}, one can expect that the necessary condition is weaker. In fact, it has been shown \cite{Khaetskii_Coish} that the smallness of the longitudinal spin decay is controlled by the parameter $\delta \! \equiv \! \mathcal{A}/\Omega\sqrt{N}$, and the condition of $\delta \! \ll \! 1$ is much less restrictive, especially for dots with many nuclear spins (i.e.~large $N$). We argue here that this condition controls the convergence of our results at the relevant time scales of interest in GaAs spin qubits (as suggested in Ref.~\onlinecite{Yao}), provided that the nuclear bath is thermal, uncorrelated, and unpolarized.
Our theory can be used to calculate decoherence under any sequence of ideal $\pi$ pulses driving the qubit (e.g.~SE corresponds to $t/2$-$\pi$-$t/2$ sequence), and we apply it to the case of Free Induction Decay (FID) and SE. For FID we obtain results which at low fields differ qualitatively from the previously obtained ones \cite{Yao,Coish_PRB08}.
For SE we identify the most relevant process contributing to its decay: the hetero-nuclear (i.e.~involving nuclei of different species) flip-flop of a pair of nuclei. At high $B$ this leads to a small oscillation superimposed on the SE signal due to the spectral diffusion, while at low $B$ it leads to a  decay of the SE in $\sim \! 0.1-1 \mu$s in gated GaAs dots. This result might  explain  the recent measurements of the SE decay on this time-scale in GaAs \cite{Petta_Science05,Koppens_PRL08}.

Most of the successful theories of spin decoherence
\cite{Witzel,Yao,Saikin_PRB07,Yang_CCE_PRB08}  involve an exponential 
resummation of the perturbation series for the decoherence
time-evolution function $W(t) \! = \! | \rho_{+-}(t) |$ 
(where $\rho_{+-}$ is the off-diagonal element of the electron spin
reduced density matrix).  We follow an approach along these lines. We take advantage of the essentially ``infinite range'' of the
hf-mediated interactions which couple effectively all the 
$N \! \sim \! 10^{5}-10^{6}$ nuclear spins in the bath (i.e.~in the
dot). 
We consider the \emph{ring diagrams},
identified as the most important contributions for long-range interactions \cite{Saikin_PRB07} in the
perturbation expansion of $W(t)$.
Corrections to these are suppressed as $1/N$.
This approach is closely related
to the $1/z$ ($z$ being the coordination number) expansion of the
partition function of the high-density Ising model \cite{Brout_PR60},
as well as recent calculations of the influence of the hf interactions on the electric dipole spin resonance in quantum dots \cite{Rashba_08}. Taking into account the ring diagrams only, we can sum all the terms in the cumulant expansion of $W(t)$.

\begin{figure}[t]
\centering
\includegraphics[width=0.9\linewidth]{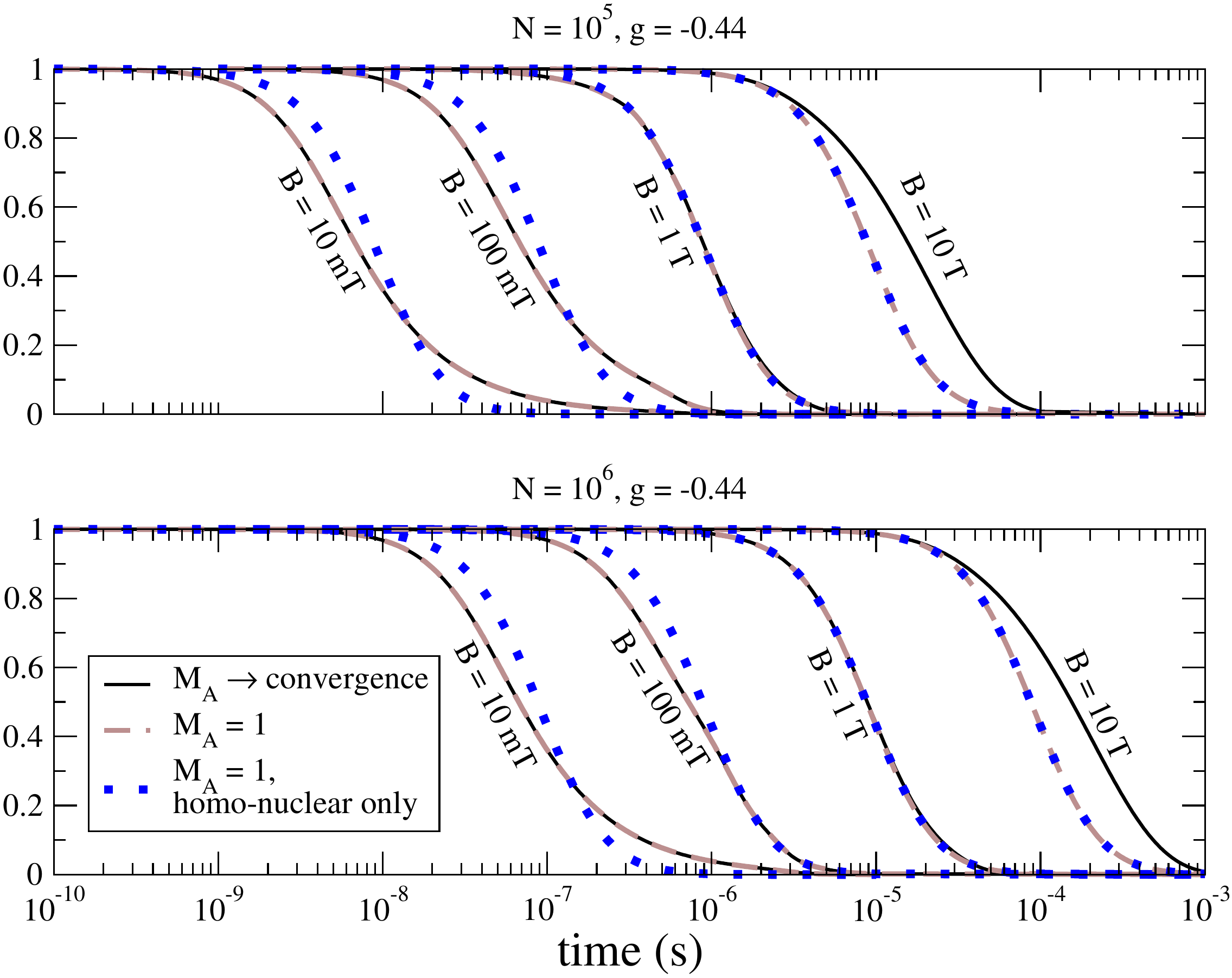}
  \caption{$W(t)$ due to hf-mediated interactions for FID calculated for two GaAs dots of different sizes and various $B$. The solid lines are the results of the full $\mathbf{T}$-matrix calculations (with convergence achieved at matrix size $M \! \approx \! 100$), dashed lines are obtained with $3 \!\times \! 3$ $\mathbf{T}$-matrix taking into account hetero-nuclear interactions, the dotted lines use Eq.~(\ref{eq:WFID_short}).
For GaAs we have used $n_{\alpha} \! = \! 0.604,0.396,1$, $\mathcal{A}_{\alpha} \! = \! 5.47, 6.99, 6.53 \cdot 10^{10}$ s$^{-1}$, and $\omega_{\alpha}/B \! = \! -6.42,-8.16,-4.58 \cdot 10^{7}$ s$^{-1}T^{-1}$ for nuclei of $^{69}$Ga, $^{71}$Ga, and $^{75}$As, respectively. The electron spin splitting is $\Omega \! = \! 0.88 \, g_{\text{eff}} B[T] \cdot 10^{11}$ s$^{-1}$ with $g_{\text{eff}} \! = \! -0.44$ in a large quantum dot. At high $B$, the actual decay will be bounded by $\sim \! 10$ $\mu$s due to spectral diffusion \cite{Yao}. } \label{fig:FID}
\end{figure}

\textit{Effective Hamiltonian}.--- The original Hamiltonian consists of Zeeman, hf, and dipolar interactions. The latter will be neglected below, since decoherence due to them can be calculated using well-tested methods \cite{Witzel}. The Zeeman energies are $ \hat{H}_{\text{Z}} \! = \!  \Omega S^{z} + \sum_{i} \omega_{\alpha[i]} J^{z}_{i}$ and hf interactions  are given by $\hat{H}_{\text{hf}}  = \sum_{i} A_{i} \mathbf{S}\cdot \mathbf{J}_{i} =  \sum_{i} A_{i}S^{z}J^{z}_{i} + \hat{V}_{\text{sf}}$, where $i$ labels the nuclear sites and $\alpha[i]$ labels the nuclear species at $i$-th site (assigned randomly to sites in case of multiple isotopes, e.g.~$^{69}$Ga and $^{71}$Ga). 
$\hat{V}_{\text{sf}} \! \sim \! S^{\pm}J^{\mp}$ is the spin-flip part of the hf interaction. 
The hf coupling $A_{i}$ is proportional to
its species' total hf energy $\mathcal{A}_{\alpha[i]}$ 
and the square of the envelope function
$f_i \! = \! |\Psi(\mathbf{r}_{i})|^{2}$: 
$A_{i} \!= \! \mathcal{A}_{\alpha[i]} f_i$ with normalization $\sum_i f_i \!=\! n_{c}$ ($n_{c}$ being the number of nuclei in the unit cell, equal to $2$ in GaAs). The number $N$ of nuclei interacting appreciably with the electron is
defined by $N\! \equiv \! \sum_{i}f_{i}/\sum_{i} f_i^2$.
For any $\alpha$, $\sum_{i \in \alpha} A^{2}_{i} \! = \!
n_{\alpha}\mathcal{A}_{\alpha}^{2}/N$ 
where $n_{\alpha}$ is the number of $\alpha$ nuclei per unit cell, and
the maximal $A_{i\in\alpha} \! \approx \! \mathcal{A}_{\alpha}/N$.
In calculations below we use a 2D Gaussian wave-function (with the parameters for GaAs given in Fig.~\ref{fig:FID}), but  the results on time-scales relevant for decay at low fields ($t \! \ll \! N/\mathcal{A}$) are determined only by $N$, not by detailed shape of $\Psi(\mathbf{r})$. 

We derive the effective pure dephasing Hamiltonian $\tilde{H}$ by
removing the $\hat{V}_{\text{sf}}$ interaction with a 
canonical transformation \cite{Yao,Coish_PRB08}. 
In the second order with respect to $\hat{V}_{\text{sf}}$ we obtain the two-spin (2s) interaction considered in Refs.~\cite{Yao,Coish_PRB08}:
$\tilde{H}^{(2)}_{2s} = 2S^{z}\sum_{i\neq j} B_{i , j}
J^{+}_{i}J^{-}_{j}$ with $B_{i , j}=A_{i} A_{j} / 4 \Omega$.
We will refer to this term as the $S^{z}$-conditioned term. 
These non-local hf-mediated flip-flop interactions are the most
important at moderate fields and yield the effective Hamiltonian
\beq
\label{eq:H2}
\tilde{H}^{(2)} = \sum_{i} \omega_{\alpha[i]} J^{z}_{i} + \sum_{i}
A_{i}S^{z}J^{z}_{i} + 2S^{z}\sum_{i\neq j} B_{i , j} J^{+}_{i}J^{-}_{j}. 
\eeq
The electron Zeeman term was discarded as being a constant of
motion; the nuclear Zeeman terms, however, are important because $B_{i, j}$
can couple nuclei of differing species (this is crucial for SE).
At higher orders we will have additional $S^{z}$-independent and multi-spin
interactions which we address near the end of this Letter.

\textit{Decoherence function}.--- We write $\tilde{H} \! = \! \HH_{0} + 2S^{z}\hat{V}_{1} + \hat{V}_{2}$,  with the flip-flop interactions divided into $S^{z}$-conditioned $\hat{V}_{1}$ and $S^{z}$-independent $\hat{V}_{2}$. We define the function $f(t;t')$ which encodes the sequence of $n$ pulses applied to the qubit \cite{Cywinski_PRB08}, $f(t;t') \! \equiv \!  \sum_{k=1}^{n} (-1)^{n+k} \Theta(t'-t_{k}) \Theta(t_{k+1}-t')$,
where $t_{k}$ with $k\! = \!1,...,n$ are the times at which the pulses are applied, and $t_{0}\! = \! 0$, $t_{n+1} \! = \! t$, with $t$ being the total evolution time. The decoherence function $W(t)$ can be written introducing the Keldysh contour \cite{Saikin_PRB07,Yang_CCE_PRB08,Lutchyn_PRB08}
\beq
W(t) = \left | \left\langle \mathcal{T}_{C} \exp \left[ -i\int_{C} [ cf(t;\tau) \mathcal{V}_{1}(\tau_{c}) +   \mathcal{V}_{2}(\tau_{c})]  d\tau_{c} \right] \right\rangle \right | \,\, . \label{eq:W_contour}
\eeq
where $\mathcal{T}_{C}$ denotes the contour-ordering of operators, $\mean{...}$ is the ensemble average, 
$\tau_{c}\! \equiv \! (\tau,c)$ is the time variable on the contour, with $c\!=\! + (-)$ on the upper (lower) branch of the contour, and $\mathcal{V}_{1,2}(\tau_{c})$ are the interactions in which the spin operators are given by
\beq
J^{\pm}_{j}(\tau_{c}) = J^{\pm}_{j} e^{ \pm i \omega_{j}\tau \pm i c \int_{0}^{\tau}f(t;t')\frac{A_{j}}{2} dt' } \equiv  J^{\pm}_{j} d^{\pm}_{j}(\tau_{c}) \,\, .
\eeq
One remark is in order for the case of FID, in which the above formulas refer to the \textit{single-spin} decoherence, defined in \cite{Yao} where it was shown that in unpolarized thermal bath one can factor out the inhomogeneous broadening from $W(t)$ (this is equivalent to FID calculation using a ``narrowed'' state with no net nuclear polarization \cite{Coish_PRB08}).

\textit{Ring diagrams}.--- We concentrate now on interactions from Eq.~(\ref{eq:H2}).
Perturbation expansion of $W(t)$ is $W \! = \! 1+ \sum_{k=1} W^{(2k)}$, with $W^{(2k)} \! \sim \! \mean{\mathcal{V}^{2k}}$.
We use here the thermal nuclear bath at temperature very high compared to the nuclear Zeeman energies, so that $\mean{...} \! \sim \! \text{Tr}\{...\}$, and inside the averages all the   $J^{+}_{i}$ operators have to be paired up with $J^{-}_{i}$ operators. Thus, $W^{(2k)}$ has the following structure:
\beq
W^{(2k)} \sim \sum_{i_{1}\neq j_{1}} ... \!\! \sum_{i_{2k}\neq j_{2k}} \mean{\mathcal{T}_{C} J^{+}_{i_{1}}J^{-}_{j_{1}} \,\, ... \,\, J^{+}_{i_{2k}}J^{-}_{j_{2k}} } \,\, ,  \label{eq:Wk}
\eeq
with at most $2k$ indices different. The full diagrammatic expansion is very cumbersome, since the commutation rule $[J^{+}_{i},J^{-}_{j}] \! = \! 2\delta_{ij} J^{z}_{i}$ does not allow the standard form of Wick's theorem to work \cite{Saikin_PRB07,Yang_CCE_PRB08}. However, for ``infinite'' range interaction (all $N$ spins coupled comparably with each other) we have $\sim\! N^{2k}$ terms in Eq.~(\ref{eq:Wk}). From the sum over $2k$ indices we now take only the terms in which none of the indices is repeated. This amounts to introducing a $1/N$ error. The simplification is tremendous: now we only have one pair of $J^{+}_{i}$ and $J^{-}_{i}$ for given $i$, and they effectively commute, as the averages containing $J^{z}_{i}$ as the sole operator of the $i$-th nuclei are zero in uncorrelated and unpolarized bath. We can then get rid of time-contour ordering, and classify all the terms in the perturbation series in terms of ring diagrams, which involve clusters of nuclei connected by two-spin interactions in a cyclical manner. Introducing the matrix $T_{ij}$ we can write the ring diagram containing $2k$ nuclei as
\beq
R_{2k}(t) =  \!\!\!\! \sum_{i_{1}\neq i_{2} \neq ... \neq i_{2k} }\!\!\!\!  T_{i_{1}i_{2}}(t) ... T_{i_{2k}i_{1}}(t) \simeq \text{Tr} \, \mathbf{T}(t)^{2k} \,\, ,  \label{eq:R_Tsum}
\eeq
where $T_{ij} \! \equiv \!  (1\!-\!\delta_{ij})\sqrt{a_{i}a_{j}}B_{ij}\! \int_{C} cf(t;\tau) d^{+}_{i}(\tau_{c}) d^{-}_{j}(\tau_{c}) d\tau_{c}$, with $a_{i}\!\!\equiv\!\!\frac{2}{3}J_{i}(J_{i}\! +\! 1)$.
With the same 1/N error, we may approximate all of the terms in the expansion of
$W(t)$ as products of $R_{2k}$ defined in Eq.~(\ref{eq:R_Tsum}), and $\ln W(t)$ is the sum of the linked terms \cite{Kubo_JPSJ62}, leading to
\beq
W(t) = \exp \Big [ \sum_{k=1}^{\infty} \frac{(-1)^{k}}{2k} R_{2k}(t) \Big ]  \,\, . \label{eq:W_R}
\eeq

Using Eqs.~(\ref{eq:R_Tsum}) and (\ref{eq:W_R}) we arrive at formula involving the eigenvalues $\lambda_{l}$ of the $\mathbf{T}$-matrix: $W(t) \! = \! \prod_{l} ( 1+ \lambda^{2}_{l} )^{-1/2}$. Since $T_{ij}$ depends only on $\omega_{i}$ and $A_{i}$, we may approximate $R_{2k}$ by considering continuous distribution $\rho(A)$ of $A_{i}$ and  then coarse-graining it. With the relevant ranges of $A_{i}$ divided into $M_{A}$ slices and with $N_{J}$ nuclear species, we have to deal with matrices of $M \times M$ size with $M \! = \! M_{A}N_{J}$. The calculation of $W(t)$ quickly converges as we increase $M$ with the coarse-graining error negligible for $t \! \ll \! NM_{A}/\mathcal{A}$.
In fact, we will show that $M\! = \! N_{J}$ (equal to $3$ in case of GaAs) gives a good approximation at experimentally relevant short times.

\textit{Free Induction Decay}.--- The main source of the FID decoherence is the interaction from Eq.~(\ref{eq:H2}) with only homo-nuclear flip-flops taken into account (as long as $\omega_{\alpha\beta} \! \equiv \! \omega_{\alpha}-\omega_{\beta} \! \gg \! \mathcal{A}/N$, which is fulfilled in GaAs with $N \! \approx \! 10^{6}$ and $B \! \approx \! 0.1$ T). Then we can consider only ring diagrams involving only the nuclei of the same species, so that $T_{ij} \! \sim \! \sin(A_{ij}t/2)/A_{ij}$, with $A_{ij} \! =\! A_{i}\!-\! A_{j}$, and 
$R_{2k} \! = \! \sum_{\alpha} R^{\alpha}_{2k}$, so that $W_{\text{FID}}(t) \! =\! \prod_{\alpha} W^{\alpha}_{\text{FID}}(t)$. 
For short times, when $t \! \ll \! N/\mathcal{A}$ (i.e.~$t\ll 20$ $\mu$s for GaAs with $N \! = \! 10^{6}$), we obtain a very simple result: $R^{\alpha}_{2k} \! = \!  (R^{\alpha}_{2})^{k} $ and $R^{\alpha}_{2} \! = \! \xi_{\alpha}t^{2}$, leading to the resummation of the exponentiated series in Eq.~(\ref{eq:W_R}) giving us
\beq
W^{\alpha}_{\text{FID}}(t)  =  \frac{1}{\sqrt{1+ \xi_{\alpha}t^{2} } }  \,\,\,\, ,  \,\,\,\, \xi_{\alpha} = a^{2}_{\alpha} \frac{n^{2}_{\alpha}\mathcal{A}^{4}_{\alpha}}{4N^{2}\Omega^{2}}  \,\, ,\label{eq:WFID_short}
\eeq
so that the characteristic decay time is $T_{\text{FID}} \! \sim \! N\Omega/\mathcal{A}^{2}$ as in Ref.~\onlinecite{Yao}.
In Fig.~\ref{fig:FID} we show that this formula describes very well the FID decoherence in GaAs for $B \! < \! 1$ T.
Let us also note that for InGaAs with $N\! = \! 10^{5}$, $g_{\text{eff}}\! =\! 0.5$, and at $B\! = \! 6$ T from Eq.~(\ref{eq:WFID_short}) we obtain an estimate $T_{\text{FID}} \! \sim \! 0.6$ $\mu$s which is close to the recently observed value of $3$ $\mu$s in an optically selected ``narrowed'' state \cite{Greilich_Science06}.
Eq.~(\ref{eq:WFID_short}) should be compared with the ``pair approximation'' result of Ref.~\onlinecite{Yao}, which in our notation reads $W^{\alpha} \! = \!  \exp ( -\frac{1}{2} R^{\alpha}_{2} )$.
This agrees with Eq.~(\ref{eq:WFID_short}) for $R_{2} \! \ll \! 1$, when both formulas give $W\! \approx \! 1 - \frac{1}{2}R_{2}$, but at times at which $R_{2}$ becomes comparable to $1$ only Eq.~(\ref{eq:WFID_short}) correctly takes into account the correlations of groups of spins larger than pairs. Interestingly, for long times ($t \! \gg \! N/\mathcal{A}$), the ring diagrams with more than two spins are suppressed by oscillations in $T_{ij}$, and the only remaining term is then $R_{2} \! \sim \! t$ at this time-scale. Thus we recover the long-time Markovian behavior discussed previously \cite{Yao,Coish_PRB08}.

\begin{figure}[t]
\centering
\includegraphics[width=0.9\linewidth]{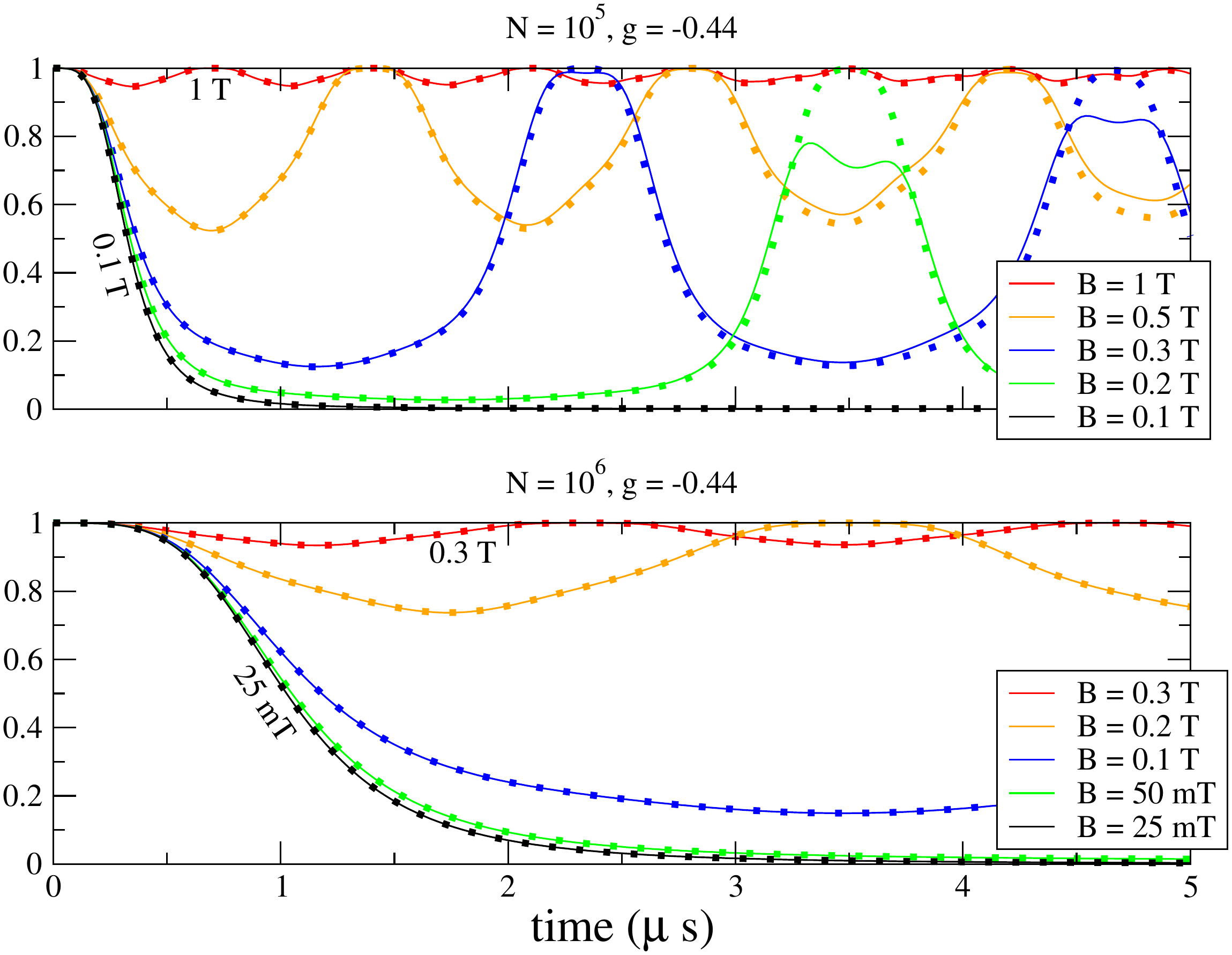}
  \caption{Spin echo decoherence $W^{\text{SE}}(t)$ in GaAs due to hetero-nuclear processes. The dots are obtained in the $\mathcal{A}/N \! \ll \! \omega_{\alpha\beta}, 1/t$ limit, when $W(t) \! = \! [1+\frac{1}{2}R_{2}(t)]^{-1}$, while the solid lines are the results of the calculation with the full $\mathbf{T}$-matrix. The differences between the two approaches on a microsecond time-scale are visible for the smaller dot (upper panel), but are negligible for the larger one (lower panel). The decay time due to spectral diffusion (calculated as in Ref.~\onlinecite{Witzel}) is $> \! 10$ $\mu$s in both cases. The GaAs parameters used are the same as in Fig.~\ref{fig:FID}.
  } \label{fig:SE}
\end{figure}

\textit{Spin Echo}.---  If we use only the $S^{z}$-conditioned interactions and allow only for homo-nuclear flip-flops, there is no SE decoherence \cite{Yao}, i.e.~$W_{\text{SE}}(t)\! = \! 1$.  The SE decay can come then either from $S^{z}$-independent interactions, or from considering the hetero-nuclear flip-flops in Eq.~(\ref{eq:H2}). The latter lead, for $t\! \ll \! N/\mathcal{A}$ and $\omega_{\alpha\beta} \! \gg \! \mathcal{A}/N$, to 
\beq
T^{\text{SE}}_{k\in \alpha, l \in \beta}  = (1\!-\!\delta_{\alpha\beta})  \frac{A_{k}A_{l}}{4\Omega} \sqrt{a_{\alpha}a_{\beta}}  \frac{8ie^{i\omega_{\alpha\beta}t/2}}{\omega_{\alpha\beta}} \sin^{2}\frac{\omega_{\alpha\beta}t}{4} \,\, ,  \label{eq:T_SE}
\eeq
The multiplicative dependence of the above matrix on $A_{k}$ and $A_{l}$ allows us to define a matrix $\tilde{T}_{\alpha\beta}$ of $N_{J} \! \times \! N_{J}$ dimension, in which $A_{k}$ is replaced by $\sqrt{n_{\alpha}/N}\mathcal{A}_{\alpha[k]}$.
Then we have $R_{2k}(t) \! = \! \sum_{l} \tilde{\lambda}^{2}_{l}$ in terms of $N_{J}$ eigenvalues of $\tilde{T}_{\alpha \beta}$. In GaAs we have $N_{J} \! = \! 3$, and we get $\tilde{\lambda}_{i} \! = \! 0,\pm \tilde{\lambda}$, so that $W_{\text{SE}}(t) \! = \! [1+ \frac{1}{2}R_{2}(t)]^{-1}$, 
where $R_{2} = 2( |\tilde{T}_{12}|^{2} + |\tilde{T}_{13}|^{2} + |\tilde{T}_{23}|^{2})$ and
\beq
 |\tilde{T}_{\alpha\beta}|^{2} =  \frac{4\mathcal{A}^{2}_{\alpha}\mathcal{A}^{2}_{\beta}}{N^{2}\Omega^{2}\omega^{2}_{\alpha\beta}} n_{\alpha}n_{\beta} a_{\alpha}a_{\beta}  \sin^{4}\frac{\omega_{\alpha\beta}t}{4} \,\, .
\eeq
The SE decoherence is shown in Fig.~\ref{fig:SE}. At high $B$ when $ |\tilde{T}_{\alpha\beta}|^{2} \! \ll \! 1$ we obtain a small oscillation which in experiments will be superimposed on the SD decay. At low $B$, which for $N\! = \! 10^{5}(10^{6})$ in GaAs corresponds to $\Omega\! < \! \Omega_{c}  \! \sim \! \sqrt{r}\mathcal{A}/\sqrt{N} \! \approx \! 250 (75)$ mT (where $r \! \sim \! 10^{3}$ is the ratio of electron $g$-factor to the typical difference of nuclear $g$-factors), the coherence decays to zero in characteristic time $T_{\text{SE}} \! \approx \! r/\Omega_{c} \approx 0.25 (0.75) \, \mu$s. This is close to $T_{\text{SE}} \approx 0.3-0.4$ $\mu$s observed in \cite{Koppens_PRL08} at $B\! \approx \! 50-70$ mT. 
The $B$ dependence of decay is visible in a narrow range of fields when the decay is substantial but not complete. 

\textit{The $S^{z}$-independent two-spin term}.--- The contribution of $\tilde{H}^{(3)}_{2s}\! \sim \! \sum_{i,j} C_{ij} J^{+}_{i}J^{-}_{j}$ with $C_{ij} \! \propto  \! A^{2}_{i}A_{j}/\Omega^{2}$  to FID and SE is calculated in the analogous way, and it turns out to be negligible for $B\! \geq \! 10$ mT and $N \! \geq \! 10^{5}$.  

\textit{Influence of multi-spin interactions}.--- The most important multi-spin terms are the ones involving $n$ nuclei in the $n$-th order in expansion of $\tilde{H}$, e.g. $\tilde{H}_{3s}^{(3)} \! \sim \! S^{z}\sum_{ijk} C_{ijk} J^{+}_{i}J^{-}_{j}J^{z}_{k}$ with $C_{ijk} \! \propto \! (\mathcal{A}/N)^{3}/\Omega^{2}$.
Applying the operator norm $|| \hat{A} || \! \equiv \! \max_{\bra{\Phi}\Phi\rangle=1} | \bra{\Phi} \hat{A} \ket{\Phi} | $ to these terms, one gets $ || \tilde{H}^{(n)}|| \!  \approx \! \mathcal{A} (\mathcal{A}/\Omega)^{n-1}$, so that it may seem  that the expansion is convergent only when $\mathcal{A}/\Omega \! < \! 1$. However, the state $\ket{\Phi}$ maximizing the norm is a highly entangled state of nuclear spins. Here we are interested in the evolution of the nuclear spin system averaged over an uncorrelated thermal ensemble, and therefore such massively entangled nuclear states do not appear in the calculations. 

We calculate the lowest order term in the expansion of $W_{3s}(t) \! \approx \! 1 + W^{(2)}_{3s} + ...$ in powers of three-spin (3s) interaction. For both FID and SE we get that for $t\ll N/\mathcal{A}$ (when $W(t)$ decays practically to zero due to the two-spin interactions)
 we have $W^{(2)}_{3s} / R_{2} \! \sim \! \delta^{2}$. Influence of each higher-order multispin term in $\tilde{H}$ comes with a higher power of the small parameter $\delta$, and 
 our neglect of these terms is a well-controlled approximation at the time-scale of interest. 
We therefore believe that the approach used in this work remains valid at relevant times down to $B\approx 10$ mT for GaAs dots with $N \! \approx \! 10^{5}-10^{6}$ nuclei.

\textit{Conclusions}.--- We have presented a theory of quantum  decoherence due to hyperfine interaction of a central spin with a thermal nuclear bath.
For both FID and SE we obtain analytical formulas for non-exponential time dependence of decoherence. 
For SE we obtain a good agreement with the measurements in gated GaAs dots, and predict the appearance of characteristic oscillations in the signal at magnetic fields slightly higher than the ones used in experiments.

We thank W.A.~Coish and A.~Shabaev for discussions. This work is supported by LPS-NSA.

\end{document}